


\font\twelverm=cmr10  scaled 1200   \font\twelvei=cmmi10  scaled 1200
\font\twelvesy=cmsy10 scaled 1200   \font\twelveex=cmex10 scaled 1200
\font\twelvebf=cmbx10 scaled 1200   \font\twelvesl=cmsl10 scaled 1200
\font\twelvett=cmtt10 scaled 1200   \font\twelveit=cmti10 scaled 1200
\font\twelvesc=cmcsc10 scaled 1200  
\skewchar\twelvei='177   \skewchar\twelvesy='60


\def\twelvepoint{\normalbaselineskip=12.4pt plus 0.1pt minus 0.1pt
  \abovedisplayskip 12.4pt plus 3pt minus 9pt
  \belowdisplayskip 12.4pt plus 3pt minus 9pt
  \abovedisplayshortskip 0pt plus 3pt
  \belowdisplayshortskip 7.2pt plus 3pt minus 4pt
  \smallskipamount=3.6pt plus1.2pt minus1.2pt
  \medskipamount=7.2pt plus2.4pt minus2.4pt
  \bigskipamount=14.4pt plus4.8pt minus4.8pt
  \def\rm{\fam0\twelverm}          \def\it{\fam\itfam\twelveit}%
  \def\sl{\fam\slfam\twelvesl}     \def\bf{\fam\bffam\twelvebf}%
  \def\mit{\fam 1}                 \def\cal{\fam 2}%
  \def\sc{\twelvesc}                   \def\tt{\twelvett}
  \textfont0=\twelverm   \scriptfont0=\tenrm   \scriptscriptfont0=\sevenrm
  \textfont1=\twelvei    \scriptfont1=\teni    \scriptscriptfont1=\seveni
  \textfont2=\twelvesy   \scriptfont2=\tensy   \scriptscriptfont2=\sevensy
  \textfont3=\twelveex   \scriptfont3=\twelveex  \scriptscriptfont3=\twelveex
  \textfont\itfam=\twelveit
  \textfont\slfam=\twelvesl
  \textfont\bffam=\twelvebf \scriptfont\bffam=\tenbf
  \scriptscriptfont\bffam=\sevenbf
  \normalbaselines\rm}



\def\beginlinemode{\endmode
  \begingroup\parskip=0pt \obeylines\def\\{\par}\def\endmode{\par\endgroup}}
\def\beginparmode{\endmode
  \begingroup \def\endmode{\par\endgroup}}
\let\endmode=\par
{\obeylines\gdef\
{}}
\def\singlespace{\baselineskip=\normalbaselineskip}

\def\oneandahalfspace{\baselineskip=\normalbaselineskip
  \multiply\baselineskip by 3 \divide\baselineskip by 2}
\def\doublespace{\baselineskip=\normalbaselineskip \multiply\baselineskip by 2}

\newcount\firstpageno
\firstpageno=2
\footline={\ifnum\pageno<\firstpageno{\hfil}\else{\hfil\twelverm\folio\hfil}%
\fi}
\def\toppageno{\global\footline={\hfil}\global\headline
  ={\ifnum\pageno<\firstpageno{\hfil}\else{\hfil\twelverm\folio\hfil}\fi}}
\let\rawfootnote=\footnote                
\def\footnote#1#2{{\rm\singlespace\parindent=0pt\parskip=0pt
  \rawfootnote{#1}{#2\hfill\vrule height 0pt depth 6pt width 0pt}}}
\def\raggedcenter{\leftskip=4em plus 12em \rightskip=\leftskip
  \parindent=0pt \parfillskip=0pt \spaceskip=.3333em \xspaceskip=.5em
  \pretolerance=9999 \tolerance=9999
  \hyphenpenalty=9999 \exhyphenpenalty=9999 }
\def\dateline{\rightline{\ifcase\month\or
  January\or February\or March\or April\or May\or June\or
  July\or August\or September\or October\or November\or December\fi
  \space\number\year}}
\def\received{\vskip 3pt plus 0.2fill
 \centerline{\sl (Received\space\ifcase\month\or
  January\or February\or March\or April\or May\or June\or
  July\or August\or September\or October\or November\or December\fi
  \qquad, \number\year)}}


\hsize=6.5truein
\hoffset=0truein
\vsize=8.9truein
\voffset=0truein
\parskip=\medskipamount
\def\\{\cr}
\twelvepoint                
\doublespace                
\overfullrule=0pt        


\newcount\timehour
\newcount\timeminute
\newcount\timehourminute
\def\daytime{\timehour=\time\divide\timehour by 60
  \timehourminute=\timehour\multiply\timehourminute by-60
  \timeminute=\time\advance\timeminute by \timehourminute
  \number\timehour:\ifnum\timeminute<10{0}\fi\number\timeminute}
\def\today{\number\day\space\ifcase\month\or Jan\or Feb\or Mar
  \or Apr\or May\or Jun\or Jul\or Aug\or Sep\or Oct\or
  Nov\or Dec\fi\space\number\year}



\def\fermino#1                 
 {\rightline{\rm FERMILAB--#1}\rightline\dateline}

\def\title                     
  {\null\vskip 3pt plus 0.2fill
   \beginlinemode \doublespace \raggedcenter \bf}

\def\author                    
  {\vskip 3pt plus 0.2fill \beginlinemode
   \doublespace \raggedcenter}

\def\affil                     
  {\vskip 3pt plus 0.1fill \beginlinemode
   \oneandahalfspace \raggedcenter \it}

\def\abstract                        
  {\vskip 3pt plus 0.3fill \centerline{\bf ABSTRACT}\smallskip
   \beginparmode\narrower\noindent}

\def\endtopmatter                
  {\endpage                        
   \body}

\def\body                        
  {\beginparmode}                

\def\head#1{                        
  \goodbreak\vskip 0.4truein        
  {\immediate\write16{#1}
   \raggedcenter {\sc #1} \par }
   \nobreak\vskip 0truein\nobreak}

\def\beneathrel#1\under#2{\mathrel{\mathop{#2}\limits_{#1}}}

\def\refto#1{$^{[#1]}$}             

\def\references                 
  {\head{References}            
   \beginparmode
   \frenchspacing \parindent=0pt    
   \parskip=0pt \everypar{\hangindent=0pt\hangafter=1}}

\gdef\refis#1{\item{#1.\ }}                        

\gdef\journal#1,#2,#3,#4.{                
    {\sl #1~}{\bf #2}, #3 (#4)}                

\def\refstylenp{                
  \gdef\refto##1{ [##1]}                        
  \gdef\refis##1{\item{[##1]\ }}                
  \gdef\journal##1,##2,##3,##4 {                
     {\sl ##1~}{\bf ##2~}(##3)~##4 }}

\def\endreferences{\body}

\def\figurecaptions                
  {\endpage
   \beginparmode
   \head{Figure Captions}
}

\def\endpage                        
  {\vfill\eject}

\def\endpaper                        
  {\endmode\vfill\supereject}


\def\ref#1{Ref.~#1}                     
\def\[#1]{[\cite{#1}]}
\def\cite#1{{#1}}
\def\(#1){(\call{#1})}
\def\call#1{{#1}}
\def\taghead#1{}
\def\frac#1#2{{#1 \over #2}}

\def\12{{1\over2}}

\def\sla{\raise.15ex\hbox{$/$}\kern-.57em}
\def\leaderfill{\leaders\hbox to 1em{\hss.\hss}\hfill}
\def\twiddle{\lower.9ex\rlap{$\kern-.1em\scriptstyle\sim$}}
\def\bigtwiddle{\lower1.ex\rlap{$\sim$}}
\def\gtwid{\mathrel{\raise.3ex\hbox{$>$\kern-.75em\lower1ex\hbox{$\sim$}}}}
\def\ltwid{\mathrel{\raise.3ex\hbox{$<$\kern-.75em\lower1ex\hbox{$\sim$}}}}
\def\square{\kern1pt\vbox{\hrule height 1.2pt\hbox{\vrule width 1.2pt\hskip 3pt
   \vbox{\vskip 6pt}\hskip 3pt\vrule width 0.6pt}\hrule height 0.6pt}\kern1pt}
\def\tdot#1{\mathord{\mathop{#1}\limits^{\kern2pt\ldots}}}

\def\pmb#1{\setbox0=\hbox{#1}%
  \kern-.025em\copy0\kern-\wd0
  \kern  .05em\copy0\kern-\wd0
  \kern-.025em\raise.0433em\box0 }



\refstylenp
\def\dalack{\footnote*{Work supported by the U.S. Department of Energy under
contract no. DOE-AC02-76-CHO-3000.}}
\fermino{PUB--92/149--T}
\title{Seesaw Neutrino Mass\\ Ratios with\\ Radiative Corrections}
\author{D.~C.~Kennedy\dalack}
\affil{Fermi National Accelerator Laboratory\\
P.O. Box 500, Batavia, Illinois 60510}
\abstract{
Unlike neutrino masses, the ratios of neutrino masses can be
predicted by up-quark seesaw models using the known quark masses and
including radiative corrections, with some restrictive
assumptions.  The uncertainties in these ratios can be reduced to
three: the type of seesaw (quadratic, linear, etc.), the top quark mass,
and the Landau-triviality value of the top quark mass.}
\endtopmatter
\body

The inconclusive but suggestive results of recent solar and atmospheric
neutrino and beta decay experiments\refto{1}
lead to the possibility of neutrino masses, which additionally
may have important application to cosmology, astrophysics and laboratory
searches for neutrino oscillations.
The most economical model of light neutrinos is the so-called ``seesaw'' of
the grand-unified type, which requires a superheavy
right-handed neutrino for each ordinary neutrino and arises naturally
in partially or completely unified theories with left-right symmetry, such
as SO(10)\refto{2,3,4}.
These grand unified seesaw models predict small
but non-zero Majorana masses for the ordinary neutrinos in terms of the
Dirac masses of the up-type quarks (u, c, t) and the superheavy
right-handed
Majorana masses.  These predictions are made uncertain, however, by the unknown
right-handed masses and by radiative corrections.  But the {\it ratios} of
neutrino masses are more definite in seesaw models, under some neccesary and
minimal assumptions (printed below in {\it italics})
about the physics underlying the seesaw\refto{5}.
The uncertainties in the mass ratios can then be narrowed to a handful.

The general tree-level form of the seesaw model mass matrix for three families
is:
$$\eqalign{
\left(\matrix{0&m_D\cr m^T_D&M_N\cr}\right),}\eqno(1)
$$\noindent
in the left- and right-handed neutrino basis, where each entry is a
3$\times$3 matrix.  We assume that the
upper left corner is zero, as a non-zero Majorana mass for left-handed
$\nu$ generally requires an SU(2)$_L$ Higgs
triplet, an unnatural addition to the Standard Model in light
of known electroweak neutral-current properties\refto{6}.
The Dirac matrix $m_D$ is both an SU(2)$_L$ and an SU(2)$_R$ doublet.  The
symmetric superheavy Majorana mass matrix $M_N$ for the right-handed neutrinos
$N$
violates lepton number, but is a Standard Model gauge singlet.  $M_N$ must be
a remnant of a broken SU(2)$_R$ or larger symmetry.
Assuming the eigenvalues
of $M_N$ are much greater than those of $m_D$, the light neutrinos acquire
a symmetric Majorana mass matrix $m_\nu =
m_DM^{-1}_Nm^T_D$ and the superheavy neutrinos a mass matrix
$M_N$ upon block diagonalization of~(1).  The
superheavy neutrinos have masses equal to the eigenvalues $(M_{N_1}, M_{N_2},
M_{N_3})$ of $M_N.$
The matrix $M_N$ can have a variety of sources\refto{3,5}.
In models with tree-level
breaking of SU(2)$_R,$ the right-handed mass requires an SU(2)$_R$ Higgs
triplet --- in SO(10) models, a Higgs {\bf 126}.  In models with
minimal Higgs content (SU(2)$_{L,R}$
singlets and doublets only, as in superstring models),
the matrix $M_N$ must arise either from loop effects\refto{3,7} or
from non-renormalizable terms, presumably induced by gravity\refto{8}.

Making predictions from the seesaw matrix $m_\nu$ requires additional
assumptions.  To obtain simple scaling dependence of light neutrinos masses
on the eigenvalues of $m_D$ requires the assumptions {\it that the matrix
$m_D$ can be freely diagonalized} and {\it that the intergenerational mixings
in $M_N$ are no larger than the ratios of eigenvalues between generations.}  We
then need to know the eigenvalues of $m_D:$ {\it here the simple grand-unified
seesaw is assumed, so that $m_D\propto m_u,$} the up-type quark mass matrix.
\footnote\dag{
The Dirac mass matrix for the charged leptons $m_l$ is proportional to the
down-type quark (d, s, b) mass matrix $m_d$ in the simplest grand unified
seesaw
models.  The matrix proportionalities of $m_D$ and $m_u,$ and $m_l$ and $m_d,$
require that each pair of
Dirac masses be generated by only or mainly one Higgs
representation.  Otherwise, a specific ansatz of Dirac masses is needed.}
For predictiveness, {\it the eigenvalues of $M_N$ are assumed proportional
to a power $p$ of the eigenvalues of $m_u.$}  The $p = 0$ and $p = 1$ cases
are the ``quadratic'' and ``linear'' seesaws, respectively, because of the
dependence of $m_{\nu ,i}$ on $m_{u,i}$\refto{4,5}.
\footnote\ddag{
If the eigenvalues of $M_N$ increase no more than linearly with the hierarchy
of eigenvalues in $m_u$ ($p\le 1$ for a simple power law), and $m_l\propto
m_d,$ then, additionally, the neutrino mixing matrix is identical to
quark CKM mixing matrix, at least at the putative unification scale.  For
reasonable values of the top quark mass, this equality approximately
holds at low energies\refto{4}.}

The family kinship of quarks and leptons in order of ascending mass is assumed;
a different kinship merely requires relabelling the neutrinos appropriately.
Forming the ratio of any two light neutrino masses,
$$\eqalign{
{m_{\nu ,i}\over m_{\nu ,j}} = {m^2_{u,i}\over m^2_{u,j}}\cdot {M^p_{N,j}\over
M^p_{N,i}},}\eqno(2)
$$\noindent
we obtain the power-law dependence of the seesaw,
with exponent $2 - p.$  Taking the ratios of neutrino masses eliminates the
overall unknown scale in $M_N.$  However, the
form~(2) requires radiative corrections to the fermion masses to arrive at
predictions.  The tree-level result~(2) is taken to be exact at
some scale $\mu = M_X$, typically the grand unification scale; the masses
$m_\nu (\mu ),$ $m_u(\mu ),$ and $M_N(\mu )$
are then run down to low energies and related to the physical masses to yield
radiatively modified seesaw predictions.  The leading logarithm approximation
is sufficient for our purposes and is evaluated here in the $\overline{MS}$
scheme.  As a number of authors have noted, much of the uncertainty in these
corrections cancels out in fermion mass ratios, if
some general conditions hold
about the physics that produces the corrections\refto{5,9}.

Corrections to the fermion masses are assumed to come from two sources,
Higgs-Yukawa couplings and gauge couplings.
{\it A generalized family symmetry is assumed for the gauge
interactions, so that, apart from
differences in mass thresholds, the gauge corrections are ``family-blind''.}
The mass matrices can then be diagonalized and corrections applied to
individual eigenvalues.
Higgs corrections to the masses are proportional to their underlying Yukawa
couplings.  For the light $\nu ,$ these are negligible, as they are for
the up-type quarks, except for the top quark.\footnote*{The large top quark
Yukawa coupling also leads to renormalization group corrections to the
first-third and second-third family CKM quark mixings.}
For the superheavy $N,$ the eigenvalues $M_{N,i}(X)$ are proportional to the
power $p$ of the eigenvalues $m_{u,i}(X).$

Considering only gauge corrections first, the $\overline{MS}$ renormalization
group equations for the fermion masses and gauge couplings $1...n...$ are
standard\refto{10}:
$$\eqalign{
{d\ln m(\mu )\over d\ln\mu} &= \sum_n b^{(n)}_m\cdot g^2_n(\mu ),\cr
{dg^2_n(\mu )\over d\ln\mu} &= -2b_n\cdot g^4_n(\mu ),}\eqno(3)
$$\noindent
with the general solution
$$\eqalign{
m(\mu )/m(\mu_0) = \prod_n \bigl\lbrack g_n(\mu )/g_n(\mu_0)
\bigr\rbrack^{-b^{(n)}_m/b_n}.}\eqno(4)
$$\noindent
The $\nu$ mass ratios at the scale
$M_X$ are the same as the physical ratios:
$$\eqalign{
m_{\nu ,i}(X)/m_{\nu ,j}(X) = m_{\nu ,i}/m_{\nu ,j} .}
\eqno(5)
$$\noindent
The equality holds because the known and unknown
gauge corrections to light neutrino masses
are due to heavy, flavor-blind interactions that begin to run only at the
$W$ boson mass, far above any neutrino mass.
The gauge corrections to the up-type quark mass ratios are substantial, because
they partly arise from QCD and because the quark masses have a large hierarchy
in the presence of massless gauge bosons.
To evaluate these corrections completely requires the assumption that {\it
there are no new particles of mass between the $Z$ boson and top quark masses
with Standard Model gauge couplings.}  {\it The gauge corrections require
the top quark mass to logarithmic accuracy, which we take from the
best neutral-current data to be} $m_t =$ 160 GeV\refto{6}.
(Powers of the top quark
mass are left explicit.)  Apart from differences in mass thresholds, the
gauge corrections from QCD,
QED and the hypercharge U(1)$_Y$ are the same for all up-type quarks.  The
weak isospin
SU(2)$_L$ corrections to the quark masses are zero, since these masses are
of the Dirac type, mixing left- and right-handed fields.  Corrections due to
new gauge couplings would begin at scales above $m_t$ and would cancel in the
ratios.  With $\kappa$ = 1 GeV and taking $m_u(\kappa )$ = 5 MeV,
$m_c(\kappa )$ = 1.35 GeV\refto{10}, and $m_t$ as free
if it occurs as a power,
$$\eqalign{
m_c(X)/m_u(X) &= m_c(\kappa )/m_u(\kappa ) = 270\cr
m_t(X)/m_c(X) &= (1.90)m_t/m_c(\kappa ) = 140(m_t/100 {\rm GeV})\cr
m_t(X)/m_u(X) &= (1.90)m_t/m_u(\kappa ) = 38000(m_t/100 {\rm GeV})
.}\eqno(6)
$$\noindent
The top quark mass is defined by $m_t = m_t(m_t).$  Since $M_{N,i}(X)\propto
m^p_{u,i}(X),$ the gauge corrections to $M_N$ are accounted for in the gauge
corrections to $m_u(X).$
{\it Any corrections to $M_N$ due to new gauge interactions
either cancel in the ratios or are assumed to be weakly coupled and thus
small.}

The other set of corrections are due to the fermions' couplings to the Higgs
sector.  The Yukawa couplings and fermion masses are simultaneously diagonal.
In the neutrino mass ratios, under our assumptions,
only the Yukawa coupling to the top quark is important.  The renormalization
group equation for the top quark mass is modified from~(3) to
$$\eqalign{
{d\ln m_t(\mu )\over d\ln\mu} = \sum_n b^{(n)}_m\cdot g^2_n(\mu ) +
b^H_m\cdot [m_t(\mu )/M_W]^2,}\eqno(7)
$$\noindent
where the factor $b^H_m$ depends on the Higgs sector.  The
solution to~(7) can be written as $m_t(\mu ) = f(\mu )\cdot m_t(\mu )_0,$ where
$m_t(\mu )_0$ is the solution to~(3).  Taking $f(m_t)$ = 1,
$$\eqalign{
1 - {1\over f^2(X)} = 2b^H_m\int^{M_X}_{m_t}{d\mu\over\mu}\cdot
{m^2_t(\mu )_0\over M^2_W}.}\eqno(8)
$$\noindent
The numerical evaluation of $f(X)$ requires the function $m_t(\mu )_0$ over
the full range from the top quark mass to unification.  However,
our ignorance of this function and of the Higgs sector can be
collapsed into a single number, the Landau-triviality value of the top quark
mass, $m_{tL}.$  This is the top quark mass for which, with a fixed $M_X,$
the right-hand side of~(8) is unity and $f(X)$ diverges.  That is,
$f(\mu )$ diverges before $\mu$ reaches $M_X,$ if $m_t$ exceeds $m_{tL}.$
The triviality value $m_{tL}$ is the upper limit of the top quark mass:
$$\eqalign{
{1\over m^2_{tL}} = 2b^H_m\int^{M_X}_{m_t}{d\mu\over\mu}\cdot{m^2_t(\mu
)_0\over
M^2_Wm^2_t},}\eqno(9)
$$\noindent
with the presence of the unknown $m_t$ as the lower bound inducing only a small
logarithmic error.  (The r.h.s. of (9) contains no powers of the top quark
mass.)  Then
$$\eqalign{
f^2(X) = {1\over 1 - m^2_t/m^2_{tL}}.}\eqno(10)
$$\noindent
For example, in the minimal Standard Model, with $M_X = M_{Pl}\simeq$
1$\times$10$^{19}$ GeV, $m_{tL}\simeq$
760 GeV; in the supersymmetric (SUSY) case, with the same $M_X,$ $m_{tL}
\simeq$ 190 GeV.  Of particular interest because of its successful
prediction of the weak mixing angle, the SUSY SU(5) grand unified model
yields $m_{tL}\simeq$ 180 GeV, with $M_X\simeq$ 2$\times$10$^{16}$ GeV.
The non-SUSY SO(10) model, breaking through an intermediate left-right
model, gives $m_{tL}\simeq$ 380 GeV\refto{4,6}.

With the aforementioned assumptions, the final mass ratios for the light
neutrinos are
$$\eqalign{
m_{\nu_\mu}/m_{\nu_e} &= (270)^{2-p}\cr
m_{\nu_\tau}/m_{\nu_\mu} &= {1\over (1 - m^2_t/m^2_{tL})^{1-p/2}}\cdot
\lbrack 140\cdot m_t/100 {\rm GeV}\rbrack^{2-p}\cr
m_{\nu_\tau}/m_{\nu_e} &= {1\over (1 - m^2_t/m^2_{tL})^{1-p/2}}\cdot
\lbrack 38000\cdot m_t/100 {\rm GeV}\rbrack^{2-p}.}\eqno(11)
$$\noindent
For a given $\nu_e$ or $\nu_\mu$ mass, the $\nu_\tau$ mass can be sensitive
to the top quark mass beyond the naive seesaw dependence, because of the
triviality factor.

It would be interesting to check how varying these assumptions changes the
neutrino
mass ratios.  Unfortunately, most of the assumptions cannot be changed without
losing predictiveness.  The flavor-blindness of the gauge interactions is
especially crucial.
However, switching to a leptonic seesaw, with
$m_D\propto m_l,$ does lead to
predictive neutrino mass ratios, {\it if the eigenvalues}
$M_{N,i}(X)\propto m^p_{l,i}(X)$ {\it and all
neutrinos and charged leptons are subject only to family-blind, weakly-coupled
gauge interactions.}  Then
$$\eqalign{
m_{\nu ,i}/m_{\nu ,j} = \bigl\lbrack m_{l,i}/m_{l,j}\bigr\rbrack^{2-p}
}\eqno(12)
$$\noindent
is a good approximation.

\head{Acknowledgment}

The author wishes to thank Paul Langacker of the University of Pennsylvania
for an earlier collaboration on the seesaw\refto{4} on which this letter is
based, Miriam Leuer and Yosef Nir of the Weizmann Institute of Science for
helpful discussions, and Carl Albright of Fermilab for informative comments.

\references

\refis{1}
The most recent results for the solar, atmospheric, and Simpson 17 keV
neutrinos may be found in {\it Proc. XV$^{th}$ Conf. Neutrino Phys. \&
Astrophys. (NEUTRINO '92),} Granada, Spain (June 1992).

\refis{2}
M.~Gell-Mann {\it et al.,} in {\it Supergravity,} eds. F. van Nieuwenhuizen,
D.~Freedman (North-Holland, Amsterdam, 1979) p.~315;
T.~Yanagida, {\it Prog. Theo. Phys.} {\bf B135} (1978) 66;
S.~Weinberg, {\it Phys. Rev. Lett.} {\bf 43} (1979) 1566.

\refis{3}
P.~Langacker, {\it Phys. Rep.} {\bf 72} (1981) 185; in {\it TASI 1990,} eds.
M.~Cvetic, P.~Langacker (World Scientific, Singapore, 1991) p.~863.

\refis{4}
S.~Bludman, D.~Kennedy, P.~Langacker {\it Nucl. Phys.} {\bf B374} (1992) 373;
{\it Phys. Rev.} {\bf D45} (1992) 1810.

\refis{5}
H.~Harari, Y.~Nir, {\it Nucl. Phys.} {\bf B292} (1987) 251.

\refis{6}
P.~Langacker, M.~Luo, {\it Phys. Rev.} {\bf D44} (1991) 817;
P.~Langacker, U. Pennsylvania preprint UPR-0492-T (1992) and
private communication.

\refis{7}
E.~Witten, {\it Phys. Lett.} {\bf 91B} (1980) 81.

\refis{8}
M.~Cvetic, P.~Langacker, U. Pennsylvania preprint UPR-0505-T (1992).

\refis{9}
A.~Buras, J.~Ellis, M.~Gaillard, D.~Nanopoulos, {\it Nucl. Phys.} {\bf B135}
(1978) 66; T.-P.~Cheng, L.-F.~Li, {\it Gauge Theory of Elementary Particle
Physics} (Oxford University Press, New York, 1984) ch.~14.

\refis{10}
J.~Gasser, H.~Leutwyler, {\it Phys. Rep.} {\bf 87} (1982) 77.

\endreferences

\endpaper
\end